\documentclass[12pt]{iopart}

\usepackage{graphicx}

\begin{document}

\title{Efficient coherent internal state transfer in trapped ions using Stimulated Raman Adiabatic Passage}

\author{Jens L. S\o rensen$^*$, Ditte M\o ller, Theis Iversen, Jakob B. Thomsen, Frank Jensen, Peter Staanum, Dirk Voigt, \\ Michael Drewsen}

\address{QUANTOP - Danish National Research Foundation: Center for Quantum Optics,\\
Department of Physics and Astronomy, University of Aarhus, DK-8000, Denmark.}

\ead{jls@phys.au.dk}

\begin{abstract}
We demonstrate experimentally how the process of Stimulated Raman Adiabatic Passage (STIRAP) can be utilized for efficient coherent internal state transfer in single trapped and laser-cooled $^{40}$Ca$^+$ ions. The transfer from the D$_{3/2}$ to the D$_{5/2}$ state, is detected by a fluorescence measurement revealing the population not transfered to the D$_{5/2}$ state. A coherent population transfer efficiency at the level of 95 \% in a setup allowing for the internal state detection of individual ions in a string has been obtained. 

\end{abstract}

\maketitle

In many fields of physics, coherent transfer of population from one specific internal state to another in atoms or molecules is desirable. Notable examples are atom clocks and interferometers\cite{chu02,kasevich02} as well as transitions between atomic and molecular Bose Einstein condensates\cite{donley02,drummond02}. In quantum information processing, coherent transfer can be used to momentarily shelve an atom in a state different from one of the qubit states in connection with qubit gate operations \cite{cirac95,kaler03}, or may be used as part of a qubit readout procedure\cite{staanum04,moller06}. For such applications, the transfer has to be nearly perfect. High fidelity transfer has previously been achieved in transfer of population between internal states of single atomic ions by applying Rabi \cite{roos99}, Raman \cite{monroe95} or composite \cite{gulde03} $\pi$-pulses. Additionally rapid adiabatic passage has proved useful for manipulating the population of individual neutral atoms \cite{khudaverdyan05} and ions \cite{wunderlich05}, however the timescales, limited by achievable Rabi frequencies, were 150 $\mu$s to a few ms in these experiments. Over such timescales extremely good control of external magnetic fields must be demonstrated. To shorten the experiments higher Rabi frequencies must be achieved, which is possible if an optical Raman transition is driven. This process is referred to as Stimulated Raman Adiabatic Passage (STIRAP)\cite{oreg84,gaubatz90,bergmann98}, and it has previously been demonstrated in experiments spanning from transitions between metastable states being part of atomic or molecular $\Lambda$-systems \cite{gaubatz90,bergmann98} via excitation of high lying electronic states in an atomic ladder system\cite{broers92,cubel05,deiglmayr06} to efficient atomic beam deflection\cite{lawall94,goldner94,weitz94}. However, so far all these experiments have involved ensembles of atoms or molecules. 

Here, we report on an efficient STIRAP process between the 3D$_{3/2}$ and 3D$_{5/2}$ metastable states in single laser-cooled $^{40}$Ca$^+$ ions. This experiment demonstrates efficient STIRAP transfer in single quantum systems. The success of the presented STIRAP process points to a diversity of single trapped ion manipulation experiments, including robust entanglement schemes\cite{unanyan01}, molecular state preparation \cite{vogelius05}, coherent control of chemical reactions\cite{liu00,harich02} and efficient quantum bit readout in quantum computation\cite{kaler03,staanum04,leibfried03}. 

\begin{figure}[b]
\includegraphics[width=0.9\textwidth]{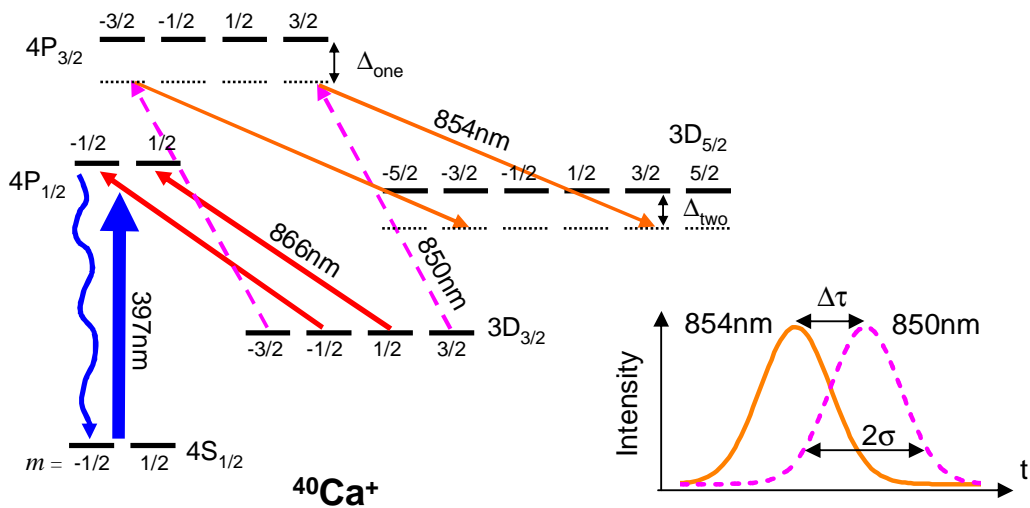}
\label{fig1}
\caption{(Color online) $^{40}$Ca$^+$ ion level scheme and laser induced transitions. The definitions of the one photon and two photon detunings are indicated as $\Delta_{\mbox{one}}$ and $\Delta_{\mbox{two}}$ respectively. (B) Sketch of the experimental setup}
\end{figure}

The relevant states and laser induced transitions in the $^{40}$Ca$^+$ ion are presented in Fig. 1, and in Fig. 2, the basics of the experimental setup is sketched. The $^{40}$Ca$^+$ ions are trapped and Doppler laser-cooled in a segmented linear Paul trap. The rf voltage ($\sim$500 V$_{pp}$, $\Omega_{rf}=2\pi\times$16.8 MHz) is delivered to the two light grey electrodes of length and thickness of 10 mm and 0.2 mm, respectively, separated by a distance of 1.4 mm. On the sectioned electrodes with section lengths: 4.5 mm, 1mm, and 4.5 mm (dark grey/blue), dc voltages of a few volts are applied to achieve axial confinement. The radial and axial trapping frequencies are typically $\sim$1.5 MHz and $\sim$0.5 MHz, respectively, allowing for confinement of few-ion strings (1-10 ions). The ions are laser-cooled using the 397 nm 4S$_{1/2} \rightarrow $4P$_{1/2}$ and the 866 nm 3D$_{3/2} \rightarrow $4P$_{1/2}$ transitions, while the STIRAP population transfer from the 3D$_{3/2}$ to the 3D$_{5/2}$ is driven by light at 850 and 854nm tuned close to the resonance frequency of the 3D$_{3/2} \rightarrow $4P$_{3/2}$ and 3D$_{5/2} \rightarrow $4P$_{3/2}$ transition, respectively. We can determine whether an ion is shelved in the 3D$_{5/2}$ state or not by exposing it simultaneously to 397 nm and 866 nm light and collecting fluorescence light at 397 nm originating from the 4S$_{1/2} \rightarrow $4P$_{1/2}$ transitions. Only when the ion is in the 4S$_{1/2}$ or the 3D$_{3/2}$ state fluorescence is detected. The internal state of the individual ions can be determined by imaging the fluorescence onto an image intensified Charge Coupled Device (CCD) camera. Additionally, a photomultiplier tube (PMT), which provide a fast and efficient way of quantifying the averaged internal state of the ensemble of ions, is used. 

\begin{figure}[b]
\includegraphics[width=0.9\textwidth]{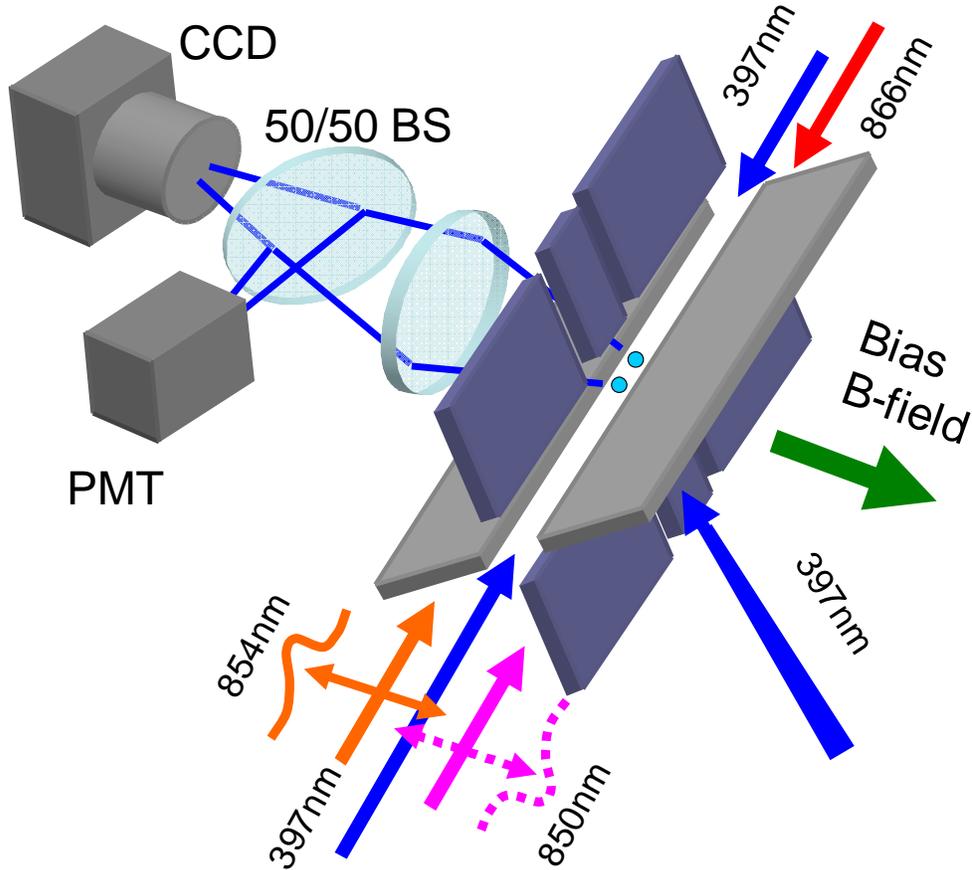}
\label{fig2}
\caption{(Color online) Sketch of the experimental setup.}
\end{figure} 
 
The 397 nm light is produced by frequency doubling the output of a Ti:Sa laser, while the remaining near infrared laser light at 850, 854 and 866 nm originates from grating stabilized diode lasers. All lasers are locked to temperature stabilized external Fabry Perot resonators with a frequency drift less than 1 MHz per hour. 

The light from the 866 nm laser is split into two beams. One, which is only present during laser cooling, passes an electro-optical phase-modulator operating at a frequency of 10 MHz for periodically scrambling the polarization of the light to avoid optical pumping into a dark state in the 3D$_{3/2}$ level\cite{siemers92}. The other beam is polarized linearly along the direction of an applied bias magnetic field of 1 Gauss, and exposes the ions for 1 ms before the STIRAP pulses are applied. This results in optical pumping into the 3D$_{3/2}(m=\pm 3/2)$ Zeeman sub states, which have the identical coupling strength with respect to the STIRAP pulses when polarized linearly along the same direction. From Raman spectroscopy we find that the optical pumping leads to a population of less than 6 \% in the 3D$_{3/2}(m=\pm 1/2)$ sublevels.

In the STIRAP experiments, the 850 and 854nm lasers are both detuned roughly $\Delta_{\mbox{one}}$=2$\pi\times$600 MHz below the 3D$_{3/2} \rightarrow $4P$_{3/2}$ and 3D$_{5/2} \rightarrow $4P$_{3/2}$ transition, respectively. The STIRAP pulses are created using the 1. order diffraction beam from acousto-optical modulators (AOM's) with controllable rf powers. The pulses generated have nearly gaussian intensity distributions, which can very well be described by $I_{850}(t)=I_{850,0} \exp[-(t-\Delta\tau/2)^2/\sigma^2]$ and $I_{854}(t)=I_{854,0} \exp[-(t+\Delta\tau/2)^2/\sigma^2]$, respectively, with $\sigma$ being the $1/e$ pulse half width, and $\Delta\tau$ being the time separation of the pulses. In the experiments, $\sigma$ and $\Delta\tau$ can both be varied from 0.5 $\mu$s to 10 $\mu$s. To avoid repumping of the population from the 3D$_{5/2}$ level to the 3D$_{3/2}$ level via the Raman resonance by residual 1. order diffracted light, an RF switch ensures that typically 10 $\mu$s after reaching the peak intensity, the RF power delivered to the AOMs is suppressed by 100dB. At a longer time scale ($\sim$ 100 $\mu$s) the STIRAP beams are blocked by a much slower mechanical shutter. In order to assure a good spatial overlap and the same linear polarization of the two STIRAP beams at the position of the ions, the light at the two wavelengths is coupled into the same single mode polarisation maintaining (PM) optical fiber using a diffraction grating. The grating also serves as a spectral filter, suppressing the spontaneous emission background of the diode lasers. For purifying the polarization of the light at the position of the ions, the collimated output beams of the fiber are send through both a $\lambda$/2- and $\lambda$/4-plate before being focused to a waist of  55 $\mu$m in the center of the trap. With peak intensities of $I_{850,0}$=42 W/cm$^2$ and $I_{854,0}$=64 W/cm$^2$, respectively, the maximum Rabi frequencies of the 3D$_{3/2} \rightarrow $4P$_{3/2}$ and 3D$_{5/2} \rightarrow $4P$_{3/2}$ transition are $\Omega_{850,0}$= 2$\pi\times$100 MHz and $\Omega_{854,0}$= 2$\pi\times$250 MHz, respectively.      
 
The bias magnetic field is actively stabilized in all three dimensions by a feedback to three pairs of current carrying coils based on the signal of a 3-axis magneto resistive sensor located about 8 cm from the trap center. The obtained stability is $\sim$1 mGauss at timescales longer than 1 second and hence much better than the absolute accuracy of the magnetic field of $\sim$20 mGauss in each dimension found from investigations of dark resonances.
Since the bias magnetic field applied along the polarization direction of the 850 nm and 854 nm beams is $\sim$ 1 Gauss, Larmor precession among the Zeeman substates of the 3D$_{3/2}$ and 3D$_{5/2}$ levels is a very small effect. Furthermore,
since the Zeeman shifts of the magnetic sublevels are much smaller than the Rabi frequencies involved, we in effect have two independent STIRAP processes coupling the $m= \pm 3/2$ sublevels of the 3D$_{3/2}$ and 3D$_{5/2}$ states with nearly the same coupling strength. This is indicated in FIG. 1. 

In the experiments, the efficiency of the STIRAP process is quantified in terms of the population transfer efficiency defined by $(\mathcal{L}-\mathcal{S})/(\mathcal{L}-\mathcal{B})$ with $\mathcal{L}$ being the scattered light level when both the 397 nm and 866 nm light is present, before the optical pumping into the 3D$_{3/2}(m=\pm 3/2)$ Zeeman substates. $\mathcal{S}$ is the light level immediately after the executing of the STIRAP sequence, and finally $\mathcal{B}$, is the background light level obtained by optically pumping all ions into the 3D$_{5/2}$ level. Each fluorescence measurement lasts 8 ms when the CCD is used and 10 ms with the PMT. The lifetime of the 3D$_{5/2}$ state is about 1.1 s \cite{staanum04a}, hence the reduction in transfer efficiency due to atomic decay is less than 1 \% in all experiments.

\begin{figure}[t]
\includegraphics[width=0.9\textwidth]{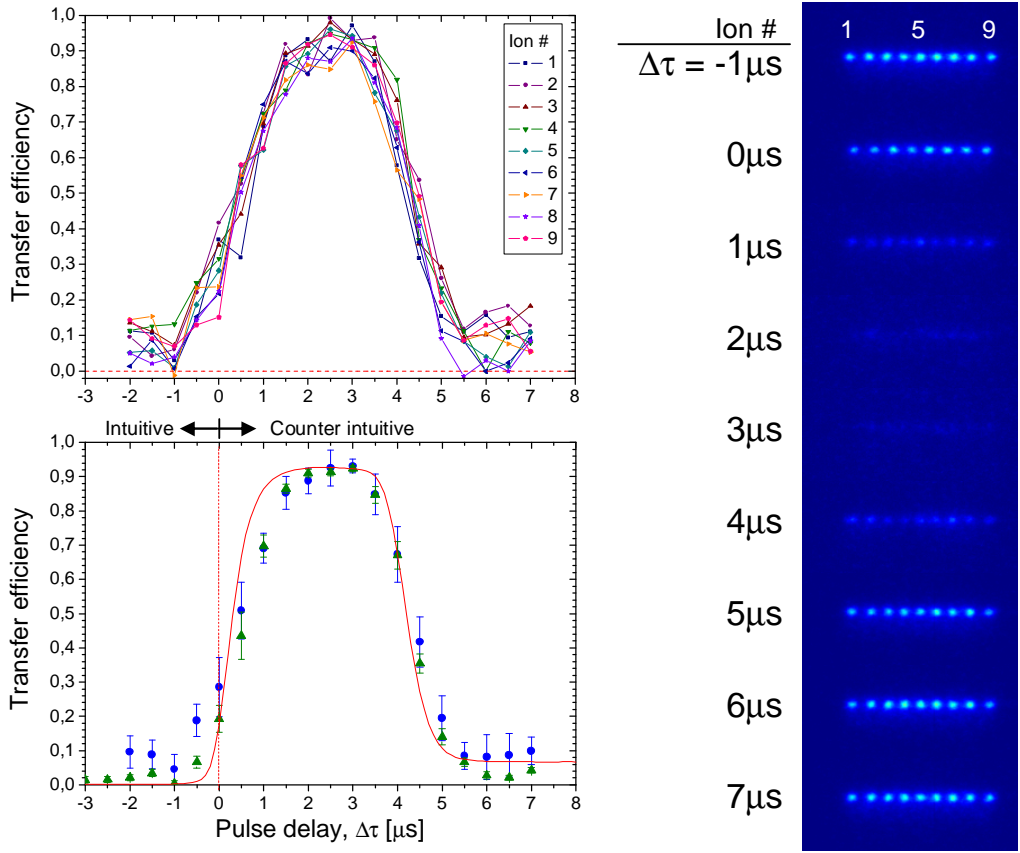}
\label{fig3}
\caption{(Color online) STIRAP transfer efficiency as function of pulse delay, $\Delta\tau$, for 9 ions on a string. Top: The STIRAP transfer efficiency data of individual ions for various pulse delays. Ion number 1 is located at the left of the pictures shown to the right. Bottom: Averaged transfer efficiency deduced from integrated CCD images (circles) and PMT counts (triangles). The full curve represents the solution to the 5 level Bloch equations, ignoring magnetic sublevels.}
\end{figure}

In Fig. 3 we present the results of an experiment where a string of 9 ions is exposed to STIRAP pulses with $\sigma$= 1.5$\mu$s, while we vary the pulse delay, $\Delta\tau$. In the experiments $\Omega_{850,0}$= 2$\pi\times$100 MHz, $\Omega_{854,0}$= 2$\pi\times$250, $\Delta_{\mbox{one}}$=2$\pi\times$600 MHz, and the two-photon 3D$_{3/2}$ - 4P$_{3/2}$ - 3D$_{5/2}$ Raman detuning is $\Delta_{\mbox{two}}$= 2$\pi\times$1 MHz. The upper graph shows the results for the individual ions on the string, while the lower graph shows the average transfer efficiency for all ions obtained by analysis of the CCD images as well as signals from the PMT. For the intuitive pulse order ($\Delta\tau<0$) we observed the smallest transfer efficiency at the level of 5 \%. Here, some oscillations are found, which we attribute to Rabi dynamics. When the STIRAP pulses perfectly overlap ($\Delta\tau=0$), the efficiency is seen to grow to about $\sim$30 \%.  For counter intuitive pulse sequences adiabatic population transfer increases the transfer efficiency until a maximum of 93$\pm$2 \% is reached for $\Delta\tau=3.0$ $\mu$s. For yet larger counter intuitive pulse delays, the efficiency drops to the level of 5 \%. This residual transfer arises from off-resonant one-photon absorption process induced by the last 850nm pulse. The full curve is obtained by solving the 5 level optical Bloch equations modelling our system in FIG. 1, ignoring magnetic sublevels. The parameters for the simulation are: $\Omega_{850,0}/2\pi=90$ MHz, $\Omega_{854,0}/2\pi=225$ MHz and $\Delta_{\mbox{one}}$=2$\pi\times$600 MHz. When the optical powers are minimum residual power levels giving rise to Rabi frequencies at 2 \% and 5 \% of the peak values, respectively, are assumed. More details about the analysis can be found in ref. \cite{moller06}. These values optimize the agreement between theory and experiment, and the discrepancy of 10 \% relative to the experimental Rabi frequencies is attributed to a nonperfect overlap between the ions and the focussed laserbeam. The role of these off-powers is discussed below.

The upper graph of Fig. 3 proves that the presented technique allows for monitoring the internal state of up to at least 9 individual ions, by integrating the fluorescence signal originating from specific regions in the trap. The analysis is based on averages of 50 STIRAP sequences, each having a duration of 40 ms and an exposure of the CCD chip in 8 ms. Here, it should be noted that the time scale of a single STIRAP sequence is much shorter than the typical time for ions to switch positions. 

The results presented in FIG. 3 do not show our highest obtained transfer efficiency. Reducing the number of ions to 2, a slightly higher transfer efficiency of $95\pm 2$ \% is measured, but in order to prove that the high transfer efficiency is not limited to a very specific spot in the trap, the results for the 9 ion string have been chosen.

In Fig. 4, the measured transfer efficiency (averaged PMT signal from 1-10 ions) as function of the two-photon detuning of the Raman transition $\Delta_{\mbox{two}}$ for pulses with $\Omega_{850,0}$=2$\pi\times$100 MHz, $\Omega_{854,0}$ = 2$\pi\times$250 MHz, $\sigma$=1.5 $\mu$s and $\Delta\tau$=2 $\mu$s. Here, as expected, a very small transfer efficiency is found for $\Delta_{\mbox{two}}>0$ due to the presence of a bright resonance\cite{bergmann98}. However, as we sweep down through the two photon resonance, we populate the dark superposition state of the 3D$_{3/2}$ and 3D$_{5/2}$ unperturbed atomic states, and hence the STIRAP effect contributes strongly\cite{bergmann98}. Due to the remaining optical power irradiating the ions after the pulse sequence, the transfer efficiency is not found to be maximum for $\Delta_{\mbox{two}}$=0, but actually for small negative two photon detunings. The relatively weak light gives rise to a narrow Raman resonance, which pump out population of the 3D$_{5/2}$ state after the STIRAP sequence when on two-photon resonance. 

Fig. 4 clearly shows that, in contrast to STIRAP experiments performed on atomic beams \cite{bergmann98}, it is not necessarily an advantage to have very high Rabi frequencies on stationary atoms due to residual optical power. In fact, much care must be taken in order to ensure spectral as well as temporal clean optical pulses in order to avoid repumping of population.

\begin{figure}[t]
\includegraphics[width=0.9\textwidth]{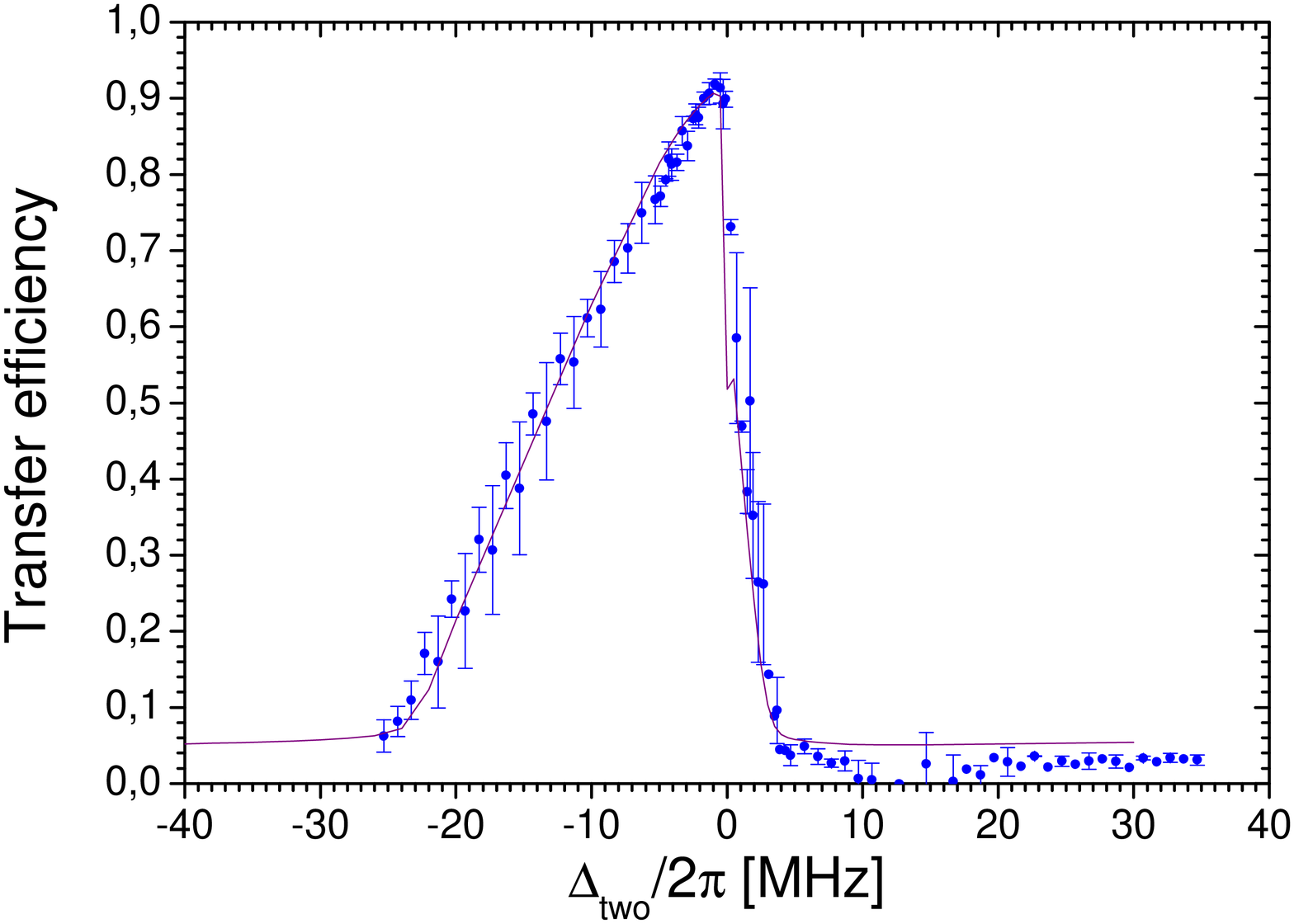}
\label{fig4}
\caption{(Color online) STIRAP transfer efficiency as function of the two-photon detuning, $\Delta_{\mbox{two}}$. Shown is an average of 5 consecutive scans with ion strings containing 1-10 ions. The data points are obtained from the average of the PMT signal. The full curve represents the solution of the optical Bloch equations involving the 5 levels shown in FIG. 1 but ignoring magnetic sublevels.}
\end{figure}

The full curve in FIG. 4 is again the solution to the 5 level Bloch equations with parameters the same as for FIG. 3.

In an attempt to measure the efficiency of our STIRAP pulses and partially circumventing the problem of residual light, we set up a pulse sequence of multiple STIRAP pulse pairs, which alternating adiabatically transfer population to 3D$_{5/2}$ state and back to the 3D$_{3/2}$ state. By using up to 7 pulse pairs, we find the final transfer efficiency to be constant at $90\pm 1$ \%, indicating that our highest transfer efficiency is indeed limited mainly by residual light on the Raman transition.

Finally, we study the role of the width of our STIRAP pulses. Here, with the Rabi frequencies available, we find the optimum transfer efficiency between $\sigma=1.5$ and 2.0 $\mu$s. For widths smaller than 1 $\mu$s, the efficiency is found to reduce rapidly due to breakdown of adiabaticity. As the width is increased to a maximum of $\sigma$=10 $\mu$s, the STIRAP efficiency drops to about 80 \%. This is attributed to decoherence between the two independent lasers involved, and it is consistent with measured laser linewidths and numerical simulations \cite{moller06}.

The Raman transition chosen in this experiment is particularly well suited for STIRAP, since the relative wavelength difference between the two lasers is only 0.5 \%. As a result, with co-propagating Raman beams, Doppler shifts cancel and we observe average transfer efficiencies exceeding 90 \%. This is the case not only when ion strings are used as a target, but also in the case of small Coulomb crystals containing 30-50 ions.

In summary, we demonstrate efficient population transfer between the two metastable levels 3D$_{3/2}$ and 3D$_{5/2}$ in cold, trapped $^{40}$Ca$^+$ ions via stimulated Raman adiabatic passage. A maximum transfer efficiency of $95\pm 2$ \% is found with gaussian intensity pulses having half width at $1/e$ height of 1.5 $\mu$s and separation 3.0 $\mu$s. At present, we believe the transfer efficiency is limited mainly by unwanted resonant Raman transitions driven by residual light, a problem we currently are working on minimizing.

\bigskip
\noindent\textbf{Acknowledgments}

This work has been supported by the Danish National Research Foundation and by the Carlsberg Foundation.
\bigskip

\noindent\textbf{References}

\bibliographystyle{unsrt}
\bibliography{jlsbib}

\end{document}